\documentstyle[12pt]{article}
\begin {document}
\begin {center}
{\Large\bf MAGNETIC VORTICES IN A GAUGED O(3) SIGMA MODEL WITH SYMMETRY
BREAKING SELF-INTERACTION}
\vskip 1in
{\bf P.Mukherjee}\footnote{e-mail:subinit@anp.saha.ernet.in}\\
{\normalsize Department of Physics\\
A.B.N.Seal College,Coochbehar\\
West Bengal,India}
\end {center}
\begin {abstract}
We consider a (2+1) dimensional nonlinear O(3) sigma model with its
U(1) subgroup gauged along with the inclusion of a self-interaction
having symmetry breaking minima.The gauge field dynamics is governed
by the Maxwell term.The model is shown to support topologically stable
purely magnetic self-dual
vortices.
\end {abstract}
\newpage
The 2+1 dimensional O(3) nonlinear sigma model has been popular
 over a long period of time[1] due to its own mathematical interest of 
providing topologically stable soliton solutions which are exactly 
integrable in the Bogomol'nyi limit [2] and also for its applications
in condensed matter physics [3,4].The topological solitons of the model
are classified according to the second homotopy of the field as a mapping
from two dimensional space to the internal field space.The solutions are
scale invariant which poses a problem for their particle interpretation
on quantisation.Numerical computation of solitonic interaction behaviour
in the model reveals the problem clearly[5].

      A particularly remarkable way of breaking the scale invariance
of the model is to partially gauge the global symmetry in the internal
space.The gauge field dynamics was initially taken to be governed by
the Maxwell term[6].A particular form of self-interaction had to be
included to satisfy self-duality conditions.This new self-duality was
later extended to models with the Chern-Simons gauge coupling[7].The
topological stability of the solitons of these models owes again to
the second homotopy of mappings from the physical to the internal space
as in the usual sigma models.These solutions also share a common feature-
they are infinitely degenerate in each topological sector.Recently
it has been shown in connection with [7] that the degeneracy of the 
topological solitons of the model is lifted by chosing a self-interaction
with symmetry breaking minima [8].The phenomenon of the lifting of the
degeneracy was associated with the novel topology introduced by the
chosen self-interaction.The topological solitons are now classified
according to the first homotopy group.
The question comes what happens when this new topology is introduced in
the model with the Maxwell term.In the present letter we address this
question.

   We consider here a (2+1) dimensional nonlinear sigma model with its
U(1) subgroup gauged by the Maxwell term.Unlike [6] the posited self-
interaction possesses symmetry breaking minima.The topological solitons
of the model are electrically neutral carrying magnetic flux which is
quantised in each topological sector.The Bogomol'nyi bounds are 
saturated.We thus observe purely magnetic self dual topological vortices.
This is in contrast with [6] where the topological solitons carry
arbitrary magnetic flux and thus do not qualify as vortices.Purely
magnetic vortices occured in a nonlinear O(3) sigma model with non-abelian
Chern-Simons term[9]and 
also
 in an
abelian model with non-minimal coupling [10].
In the latter case the solitons are non-topological in nature.
Thus it appears that the present model reveals the first occurence of
topologically stable purely magnetic vortices in the context of the O(3)
nonlinear sigma model with abelian gauge coupling..

The Lagrangian of our model is given by
\begin {equation}
{\cal L} ={1 \over 2}D_\mu {\bf \phi}\cdot D^\mu{\bf \phi}
         -{1 \over 4}F_{\mu\nu}F^{\mu\nu}+
           U({\bf \phi})
\end {equation}
Here ${\bf \phi}$ is a triplet of scalar fields constituting a vector in the
internal space with unit norm
\begin {eqnarray} 
\phi_a = {\bf n_a}\cdot {\bf { \phi}},(a=1,2,3)\\
{\bf \phi}\cdot{\bf \phi} =\phi_a\phi_a= 1
\end {eqnarray}
where ${\bf  n}_a$
constitute a basis of unit orthogonal vectors in the internal space.
We work in the Minkowskian space - time with the metric tensor
diagonal, $g_{\mu\nu} = (1,-1,-1)$.

$D_\mu {\bf \phi}$ is the covariant derivative given by
\begin {equation}
D_\mu {\bf{\phi}}=\partial_\mu {\bf{\phi}}
   + A_\mu {\bf n}_3\times{\bf{\phi}}
\end {equation}
 The SO(2)
(U(1)) subgroup is gauged by the vector potential $A_\mu$ whose dynamics is
 dictated
by the Maxwell term.The electromagnetic field tensor
\begin {equation}
F_{\mu\nu} = \partial_\mu A_\nu - \partial_\nu A_\mu
\end {equation}
The potential
\begin {equation}
U({\bf{\phi}})=-{1 \over {2}} \phi_3^2
\end {equation}
gives our chosen form of self interaction of the fields $\phi_a$.Note that 
the minima
of the potential arise when ,
\begin {equation}
\phi_3 = 0\hspace{.2cm} and \hspace{.2cm}\phi_1^2+\phi_2^2 =1
\end {equation}
Clearly (7) corresponds to the spontaneous breaking of the U(1)
symmetry of (1).

The Euler - Lagrange equations of the system (1) is derived subject to the
constraint (3) by the Lagrange multiplier technique
\begin {eqnarray}
D_\nu (D^\nu {\bf{ \phi}})& =& [D_\nu (D^\nu {\bf{ \phi}})
\cdot{\bf{ \phi}}]{\bf{ \phi}}
   -{\bf n}_3\phi_3
              +\phi_3^2{\bf \phi}\\
\partial_\nu F^{\nu\mu}& =& j^\mu
\end {eqnarray}
where 
\begin {equation}
j^\mu = -{\bf n}_3\cdot{\bf J}^\mu\hspace{.2cm} and\hspace{.2cm}
 {\bf J}^\mu ={\bf{ \phi}}\times D^\mu {\bf{ \phi}}
 \end {equation}
Using (8) we get 
\begin {equation}
D_\mu {\bf J}^\mu = ({\bf n}_3
\times {\bf {\phi}})\phi_3
\end {equation}

From (9) we find,for static configurations
\begin {equation}
\nabla^2 A^0 = -A^0(1 - \phi_3^2)
\end {equation}
From the last equation it is evident that we can chose
\begin {equation}
A^0 = 0
\end {equation}
As a consequence we find that the excitations of the model are electrically
neutral.

By inspection we can construct a conserved current
\begin {equation}
K_\mu = {1 \over {8\pi}}\epsilon_{\mu\nu\lambda}[
{\bf {\phi}}\cdot D^\nu {\bf {\phi}}
\times D^\lambda{\bf {\phi}} - F^{\nu\lambda}\phi_3]
\end {equation}
It can be shown easily that
\begin {equation}
\partial_\mu K^\mu = 0
\end {equation}
The corresponding conserved charge is
\begin {equation}
T = \int d^2x K_0
\end {equation}
Using (14) and (16) we can write

\begin {equation}
T = \int d^2x[{1 \over{8\pi}}\epsilon_{ij}{\bf {\phi}}
\cdot(\partial^i{\bf {\phi}}
\times \partial^j {\bf {\phi}})]
+{ 1 \over {4\pi}}\int_{boundary}\phi_3 A_\theta r d\theta
\end {equation}
where r,$\theta$ are polar coordinates in the physical space and $A_\theta
= {\bf e}_\theta \cdot {\bf A}$.We identify T with the topological charge
of the model.The form (17) will be useful in the following analysis.

We now construct the
 energy functional  from Schwinger's energy - momentum tensor
[11] which in the static limit becomes
\begin {equation}
E = {1 \over 2}\int d^2x[(D_i{\bf {\phi}})\cdot(D_i{\bf {\phi}})+
F_{12}^2+\phi_3^2]
\end {equation}
The energy functional  (18)
 is subject to the 
constraint (3).
For finiteness of energy we require $\bf\phi$ to go to one of the symmetry 
breaking minima given by (7) for points at infinity.The physical infinite
circle is thus mapped on the equatorial circle of the internal sphere.
This is the new topology we have talked of earlier.Note that for the usual
sigma model the topological solutions are characterised by the second
homotopy of the field as a mapping from two dimensional space to the
internal field space.Distinctively,here the topologically stable solutions
are classified according to the homotopy
\begin {equation}
\Pi_1(S_1) = Z
\end {equation}

Let us note in passing the influence of the new topology on the admissible
values of the topological charge
from (17).Wwe find that the second term on te r.h.s. vanishes 
in
 the limit
$\phi_3 \to 0$ on the boundary.The first term of the r.h.s. is known to
give the number of times the physical space  wraps the internal sphere.
Clearly,when the equatorial circle is traversed once,
the physical space is mapped on a hemisphere of the internal sphere.Thus 
the topological charge will be given by
\begin {equation}
T ={ n \over 2}
\end {equation}
where n is the number of times the equatorial circle is traversed.
Evidently half integral values of T are allowed.

Let us now
define
\begin {equation}
\psi = \phi_1 + i\phi_2
\end {equation}
Using the definition of $\psi$ we can write
\begin {equation}
D_i {\bf {\phi}}\cdot D_i{\bf{\phi}}
      = |(\partial_i + iA_i)\psi|^2 +(\partial_i\phi_3)^2
\end {equation}
From the discussions of the preceding paragraph we get
\begin {equation}
\psi \approx e^{in\theta}
\end {equation}
on the physical boundary where n has been defined in (20).
From (18) and (22) we observe that for finite energy configurations we
require 
\begin {equation}
{\bf{A}}={\bf{e_\theta}}{n\over r}
\end {equation}
on the boundary.

The asymptotic form (22) allows us to compute the magnetic flux
\begin {equation}
\Phi = \int B d^2x = \int_{boundary}A_\theta r d\theta = 2\pi n
\end {equation}
The above equation shows that the magnetic flux is quantised in each
topological sector.Thus in contrast with [6] the topologically stable solutions 
solutions of the model (1) have quantised magnetic flux.Note that
the quantisation of the magnetic flux is ensured by the novel topology (19)
the origin of which traces back to the choice of the self-interaction 
potential (6) having symmetry breaking minima.

  Our next persuit is to show that in the present model self-duality 
conditions can be obtained by satisfying Bogomol'nyi limit.For this
purpose we rearrange the energy functional as
\begin {equation}
E = {1 \over 2} \int d^{2}x[{1 \over 2}(D_i{\bf {\phi}} \pm
\epsilon_{ij}{\bf {\phi}}\times  D_{j}{\bf {\phi}})^2 +
(F^{12} \pm \phi_3)^2]
    \pm 4\pi T
  \end {equation}
  Equation (26) gives the Bogomol'nyi conditions
  \begin {eqnarray}
  D_i{\bf {\phi}}\pm  \epsilon_{ij} {\bf {\phi}}\times D_j{\bf \phi} = 0\\
  F_{12}\pm \phi_3 =0
 \end {eqnarray}
  which minimise the energy functional in a particular topological sector,
  the upper sign corresponds to +ve and the lower sign corresponds to -ve
  value of the topological charge.The equations can be handled in the usual
method[3,6] to show that the scale invariance is removed by the artifice
of the gauge-field coupling.
  
  We will now show the consistency of (27) and (28) using the well-known
  Ansatz[6,
12]
  \begin {eqnarray}
  \phi_1(r,\theta) = \sin F(r) \cos n\theta\nonumber\\
  \phi_2(r,\theta) = \sin F(r) \sin n\theta\nonumber\\
  \phi_3(r,\theta) = \cos F(r)\nonumber\\
  {\bf A}(r,\theta)= -{\bf e}_\theta {{na(r)} \over r}
  \end {eqnarray}
  From (7) we observe that we require the boundary condition
  \begin {equation}
  F(r) \to \pm {\pi \over 2}\hspace{.2cm}  
as\hspace{.2cm} r \to \infty
  \end {equation}
and equation (24) dictates that
  \begin {equation}
a(r) \to -1 \hspace{.2cm}as\hspace{.2cm} r \to \infty
  \end {equation}
Remember that equation (24) was obtained so as the solutions have finite
energy.
  Again for the fields to be well defined at the origin we require
  \begin {equation}
  F(r) \to 0 or \pi \hspace{.2cm}and \hspace{.2cm}
  a(r) \to 0\hspace{.2cm} as \hspace{.2cm}r \to 0
  \end {equation}
Substituting the Ansatz(29) into (27) and (28) we find that
  \begin {eqnarray}
  F^\prime (r) = \pm {{n(a+1)}\over r} \sin F\\
  a^\prime (r) = \mp {r \over n} \cos F
  \end {eqnarray}
  where the upper sign holds for +ve T and the lower sign corresponds to
  -ve T.Equations (33) and (34) are not exactly integrable.They may be
solved numerically subject to the appropriate boundary conditions to get
the exact profiles.

Using the Ansatz (29) we can explicitly compute the topological charge T
by performing the integration in (17).The result is
\begin {equation}
T = -{n\over 2}[cosf(\infty)-cosf(0)]
\end {equation}
So we find that according to (30) and (32) T =$\pm {n\over 2}$ which is in
agreement with our observation (20).Note that f(0)
=0
 corresponds to +ve T
and f(0) = $\pi$ corresponds to -ve T.
If we take +ve T we find F(r) bounded between 0 and 
  $\pi\over 2$
  is consistent with (30),(32) and (33).Again a(r) bounded between 0 and -1 
is consistent
  with (31),(32) and (34).Thus for +ve topological charge the ansatz (29)
 with the following
  boundary conditions
  \begin {eqnarray}
  F(0) = 0\hspace{.2cm}  a(0) = 0\nonumber\\
  F(\infty)={\pi \over 2} \hspace{.2cm}a(\infty)= -1
  \end {eqnarray}
  is consistent with the Bogomol'nyi conditions.
Similarly the consistency may be verified for -ve T.
  
     The conclusions are as follows.By chosing a self-interaction with
symmetry breaking minima in a nonlinear O(3) sigma model partially
gauged with the Maxwell gauge fields we get neutral purely magnetic
vortices.These vortices are topologically stable having quantised 
magnetic flux in each topological sector.This feature is in clear contrast
with the results of the earlier model [6] where the magnetic flux
remained arbitrary.We have ascribed the reason of this difference
to the novelty of the topology introduced in the model by the posited
self-interaction.We have also explicitly demonstrated that the model satisfies 
Bogomol'nyi limits giving self- dual vortices.These vortices share the
desirable feature of the breaking of scale-invariance and are thus
suitable for particle interpretation.The consistency of the Bogomol'nyi
conditions have been discussed subject to the appropriate boundary
conditions.


\newpage
\begin{thebibliography}{99}
\bibitem [1] {kn:xx}  A.A. Belavin and A.M. Polyakov,JETP Lett.22(1975) 245.
\bibitem [2] {kn:xx} E.B. Bogomol'nyi,Sov. J. Nucl. Phys.24(1976)449.
\bibitem [3] {kn:xx}  R.Rajaraman, Solitons and Instantons
  (North Holland Publishing Company).
\bibitem [4] {kn:xx}  W.J.Zakrzewski,Low dimensional sigma models
(A.Hilger,Bristol,1989).
\bibitem [5] {kn:xx} R.A.Leese,M.Peyrard and W.J.Zakrzewski,
Nonlinearity 3(1990)387.
\bibitem [6] {kn:xx} B.J.Schroers,Phys.Lett.B 356(1995)291.
\bibitem [7] {kn:xx} P.K.Ghosh and S.K.Ghosh,Phys.Lett.B366(1996)199.
\bibitem [8] {kn:xx} P.Mukherjee,On the question of degeneracy of topological
solitons in a gauged O(3) nonlinear sigma model with Chern-Simons term
(to be published in Phys.Lett.B).
\bibitem [9] {kn:xx} G.Nardelli,Phys.Rev.Lett.73(1994)2524.
\bibitem [10] {kn:xx} P.K.Ghosh,Phys.Lett.B 381(1996)237.
\bibitem [11] {kn:xx} J.Schwinger,Phys.Rev.127(1962)324.
\bibitem [12] {kn:xx} Y.S.Yu and A.Zee,Phys.Lett. B147(1984) 325.
\end{thebibliography}
 \end {document}